# Computer Simulation of Atoms Nuclei Structure Using Information Coefficients of Proportionality

Mikhail M. Labushev

*Department of Ore Deposits Geology and Prospecting, Institute of Mining, Geology and Geotechnology, Siberian Federal University, Russian Federation, mlabushev@yandex.ru*

The research of the proportionality of atomic weights of chemical elements [1] made it possible to obtain 3 x 3 matrices for the calculation of information coefficients of proportionality Ip that can be used for 3D modeling of the structure of the nucleus of an atom. The elements of the matrices are integers from 0 to 8, in a sum equal to 9, which are not arranged in a column or a row. All zero elements of matrices are excluded from the Ip calculation.

By computer simulation it was determined that there exist only 120 various Ip values from 0 to 1.098612, that can be calculated on the basis of the matrices (Fig.1).

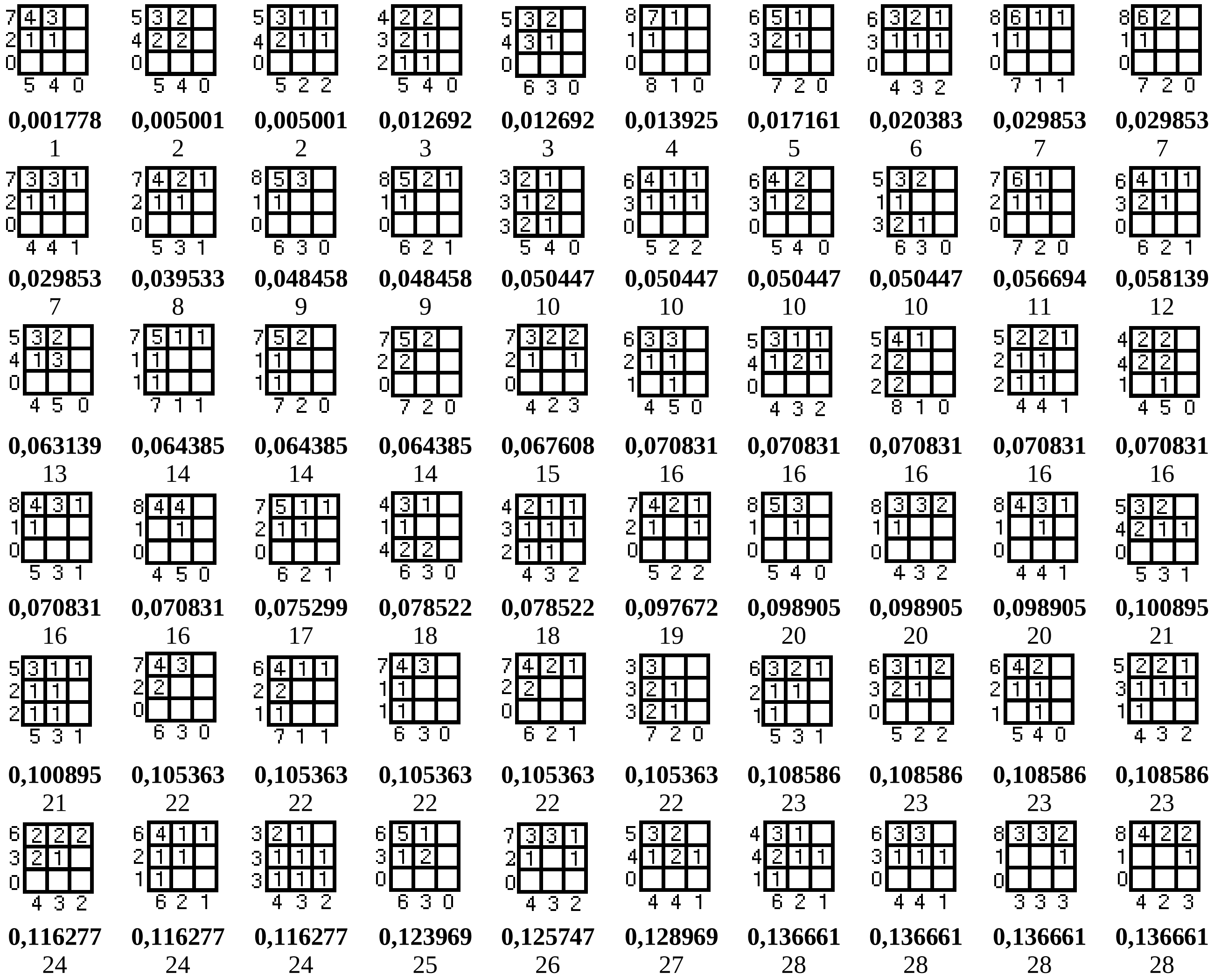

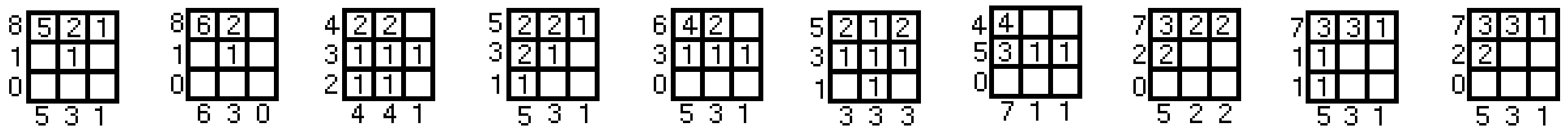

| | | | | | | | | | |
|---|---|---|---|---|---|---|---|---|---|
| **0,136661** | **0,136661** | **0,136661** | **0,13865** | **0,146341** | **0,146341** | **0,155811** | **0,155811** | **0,155811** | **0,155811** |
| 28 | 28 | 28 | 29 | 30 | 30 | 31 | 31 | 31 | 31 |

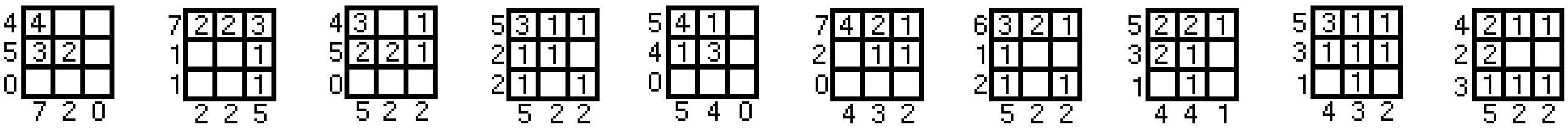

| | | | | | | | | | |
|---|---|---|---|---|---|---|---|---|---|
| **0,155811** | **0,155811** | **0,159033** | **0,159033** | **0,159033** | **0,163502** | **0,166725** | **0,166725** | **0,166725** | **0,166725** |
| 31 | 31 | 32 | 32 | 32 | 33 | 34 | 34 | 34 | 34 |

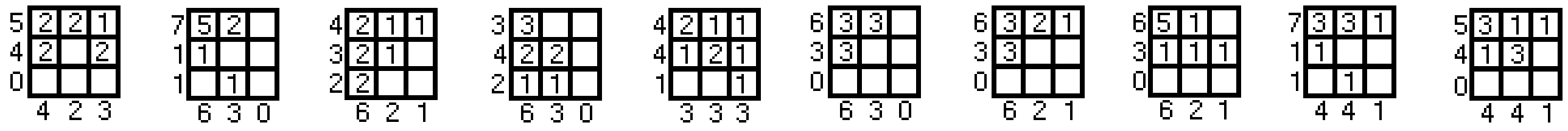

| | | | | | | | | | |
|---|---|---|---|---|---|---|---|---|---|
| **0,166725** | **0,171193** | **0,174416** | **0,174416** | **0,174416** | **0,174416** | **0,174416** | **0,182107** | **0,183885** | **0,187108** |
| 34 | 35 | 36 | 36 | 36 | 36 | 36 | 37 | 38 | 39 |

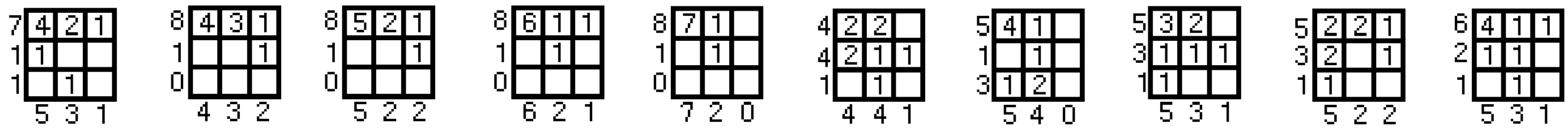

| | | | | | | | | | |
|---|---|---|---|---|---|---|---|---|---|
| **0,193566** | **0,194799** | **0,194799** | **0,194799** | **0,194799** | **0,194799** | **0,196789** | **0,196789** | **0,196789** | **0,20448** |
| 40 | 41 | 41 | 41 | 41 | 41 | 42 | 42 | 42 | 43 |

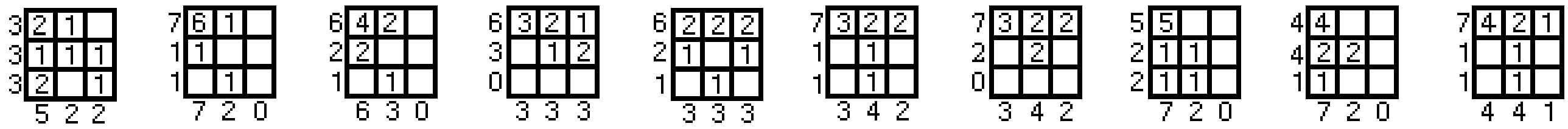

| | | | | | | | | | |
|---|---|---|---|---|---|---|---|---|---|
| **0,20448** | **0,210727** | **0,212171** | **0,212171** | **0,212171** | **0,221641** | **0,221641** | **0,221641** | **0,221641** | **0,221641** |
| 43 | 44 | 45 | 45 | 45 | 46 | 46 | 46 | 46 | 46 |

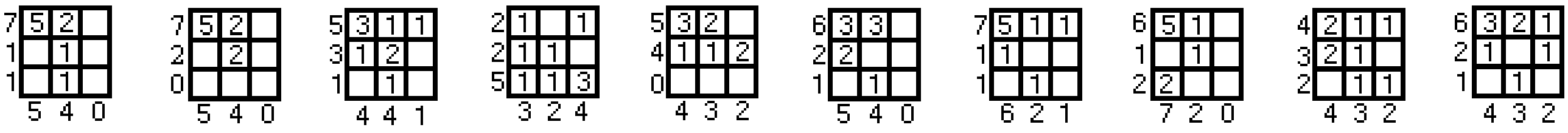

| | | | | | | | | | |
|---|---|---|---|---|---|---|---|---|---|
| **0,221641** | **0,221641** | **0,224863** | **0,224863** | **0,224863** | **0,224863** | **0,229332** | **0,229332** | **0,232555** | **0,232555** |
| 46 | 46 | 47 | 47 | 47 | 47 | 48 | 48 | 49 | 49 |

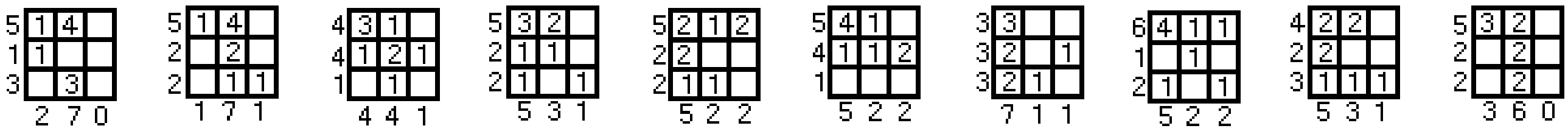

| | | | | | | | | | |
|---|---|---|---|---|---|---|---|---|---|
| **0,251705** | **0,251705** | **0,252938** | **0,254928** | **0,254928** | **0,254928** | **0,259396** | **0,262619** | **0,262619** | **0,262619** |
| 50 | 50 | 51 | 52 | 52 | 52 | 53 | 54 | 54 | 54 |

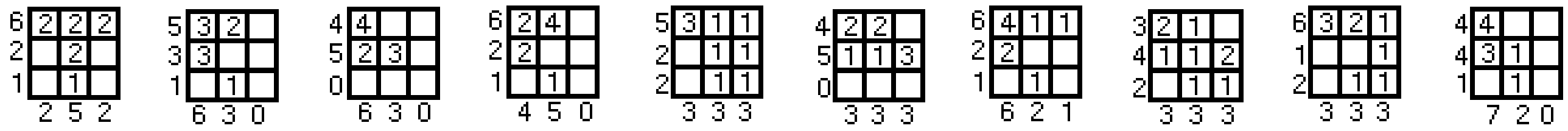

| | | | | | | | | | |
|---|---|---|---|---|---|---|---|---|---|
| **0,262619** | **0,262619** | **0,262619** | **0,262619** | **0,262619** | **0,262619** | **0,27031** | **0,27031** | **0,27031** | **0,279779** |
| 54 | 54 | 54 | 54 | 54 | 54 | 55 | 55 | 55 | 56 |

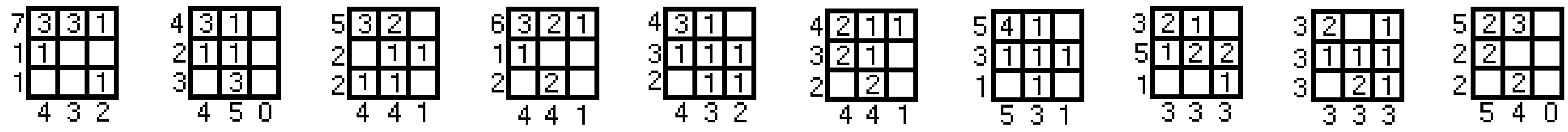

| | | | | | | | | | |
|---|---|---|---|---|---|---|---|---|---|
| **0,279779** | **0,283002** | **0,283002** | **0,290693** | **0,290693** | **0,290693** | **0,292683** | **0,300374** | **0,308065** | **0,313066** |
| 56 | 57 | 57 | 58 | 58 | 58 | 59 | 60 | 61 | 62 |

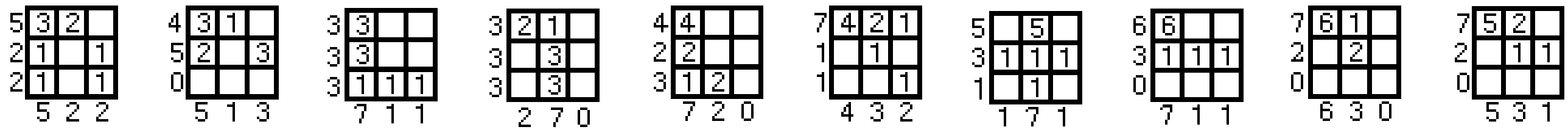

| | | | | | | | | | |
|---|---|---|---|---|---|---|---|---|---|
| **0,313066** | **0,313066** | **0,317535** | **0,317535** | **0,317535** | **0,317535** | **0,317535** | **0,317535** | **0,317535** | **0,317535** |
| 62 | 62 | 63 | 63 | 63 | 63 | 63 | 63 | 63 | 63 |

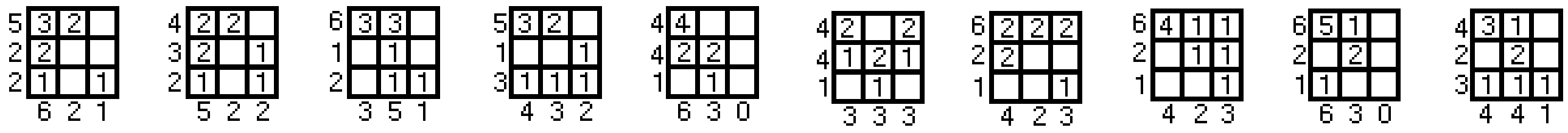

| | | | | | | | | | |
|---|---|---|---|---|---|---|---|---|---|
| **0,320757** | **0,320757** | **0,320757** | **0,320757** | **0,328449** | **0,328449** | **0,328449** | **0,328449** | **0,33614** | **0,348832** |
| 64 | 64 | 64 | 64 | 65 | 65 | 65 | 65 | 66 | 67 |

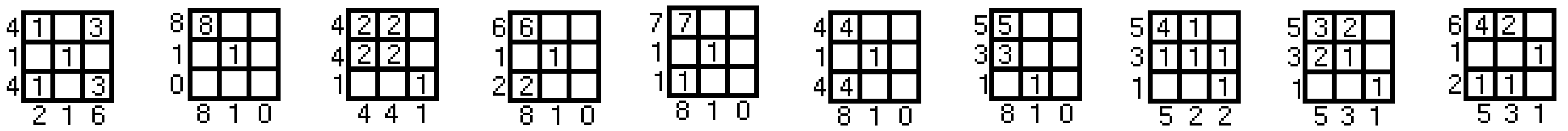

| | | | | | | | | | |
|---|---|---|---|---|---|---|---|---|---|
| **0,348832** | **0,348832** | **0,348832** | **0,348832** | **0,348832** | **0,348832** | **0,348832** | **0,350822** | **0,350822** | **0,358513** |
| 67 | 67 | 67 | 67 | 67 | 67 | 67 | 68 | 68 | 69 |

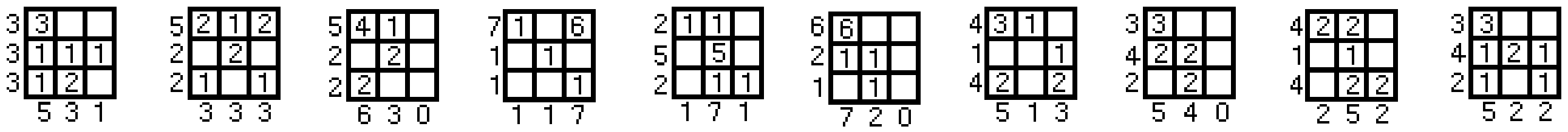

| | | | | | | | | | |
|---|---|---|---|---|---|---|---|---|---|
| **0,358513** | **0,358513** | **0,358513** | **0,36476** | **0,375673** | **0,375673** | **0,378896** | **0,378896** | **0,378896** | **0,378896** |
| 69 | 69 | 69 | 70 | 71 | 71 | 72 | 72 | 72 | 72 |

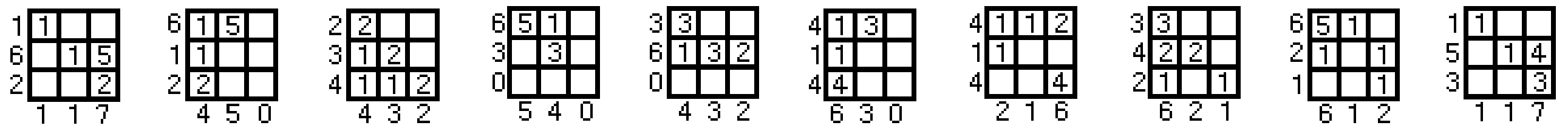

| | | | | | | | | | |
|---|---|---|---|---|---|---|---|---|---|
| **0,383365** | **0,386587** | **0,386587** | **0,386587** | **0,386587** | **0,386587** | **0,386587** | **0,386587** | **0,394279** | **0,405738** |
| 73 | 74 | 74 | 74 | 74 | 74 | 74 | 74 | 75 | 76 |

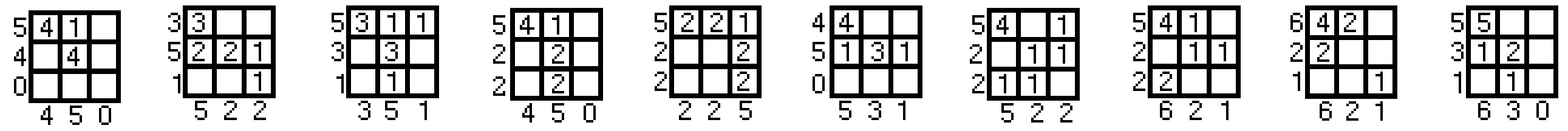

| | | | | | | | | | |
|---|---|---|---|---|---|---|---|---|---|
| **0,40896** | **0,40896** | **0,40896** | **0,40896** | **0,40896** | **0,40896** | **0,40896** | **0,416652** | **0,424343** | **0,424343** |
| 77 | 77 | 77 | 77 | 77 | 77 | 77 | 78 | 79 | 79 |

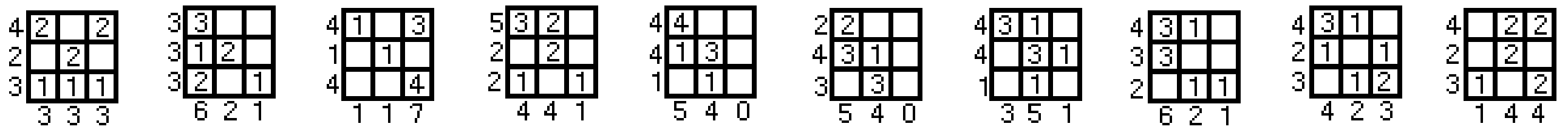

| | | | | | | | | | |
|---|---|---|---|---|---|---|---|---|---|
| **0,424343** | **0,424343** | **0,433812** | **0,437035** | **0,437035** | **0,437035** | **0,437035** | **0,444726** | **0,444726** | **0,444726** |
| 79 | 79 | 80 | 81 | 81 | 81 | 81 | 82 | 82 | 82 |

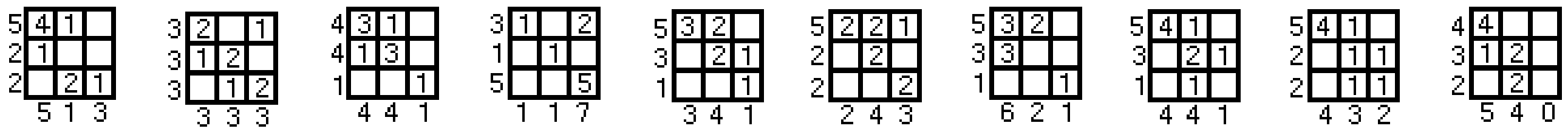

| | | | | | | | | | |
|---|---|---|---|---|---|---|---|---|---|
| **0,446716** | **0,462098** | **0,465109** | **0,471568** | **0,47479** | **0,47479** | **0,47479** | **0,47479** | **0,47479** | **0,47479** |
| 83 | 84 | 85 | 86 | 87 | 87 | 87 | 87 | 87 | 87 |

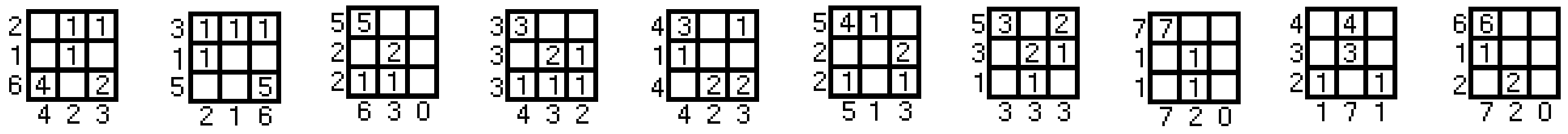

| | | | | | | | | | |
|---|---|---|---|---|---|---|---|---|---|
| **0,482481** | **0,482481** | **0,482481** | **0,482481** | **0,502865** | **0,504854** | **0,512546** | **0,529706** | **0,529706** | **0,529706** |
| 88 | 88 | 88 | 88 | 89 | 90 | 91 | 92 | 92 | 92 |

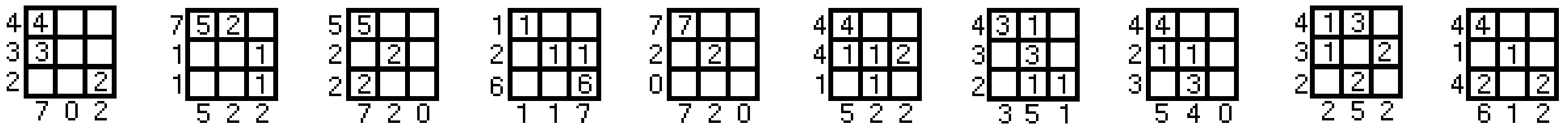

| | | | | | | | | | |
|---|---|---|---|---|---|---|---|---|---|
| **0,529706** | **0,529706** | **0,529706** | **0,529706** | **0,529706** | **0,532929** | **0,532929** | **0,532929** | **0,532929** | **0,54062** |
| 92 | 92 | 92 | 92 | 92 | 93 | 93 | 93 | 93 | 94 |

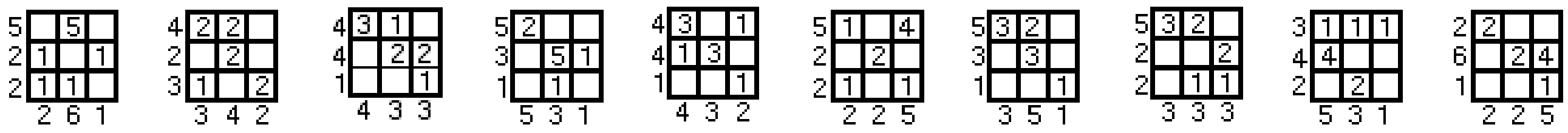

| | | | | | | | | | |
|---|---|---|---|---|---|---|---|---|---|
| **0,54062** | **0,54062** | **0,54062** | **0,548311** | **0,561003** | **0,562993** | **0,562993** | **0,570684** | **0,570684** | **0,570684** |
| 94 | 94 | 94 | 95 | 96 | 97 | 97 | 98 | 98 | 98 |

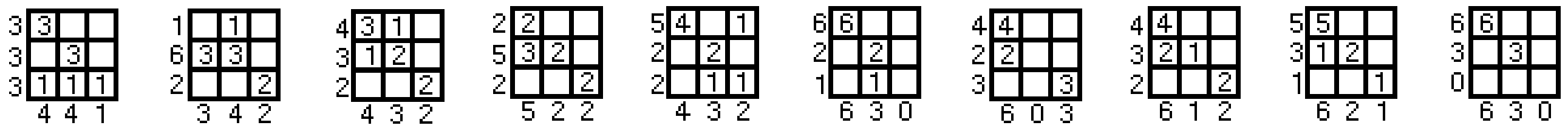

| | | | | | | | | | |
|---|---|---|---|---|---|---|---|---|---|
| **0,598759** | **0,598759** | **0,598759** | **0,621132** | **0,628823** | **0,636514** | **0,636514** | **0,636514** | **0,636514** | **0,636514** |
| 99 | 99 | 99 | 100 | 101 | 102 | 102 | 102 | 102 | 102 |

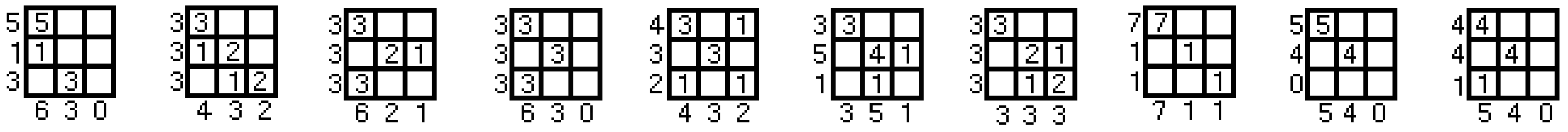

| | | | | | | | | | |
|---|---|---|---|---|---|---|---|---|---|
| **0,636514** | **0,636514** | **0,636514** | **0,636514** | **0,656898** | **0,658887** | **0,67427** | **0,683739** | **0,686962** | **0,686962** |
| 102 | 102 | 102 | 102 | 103 | 104 | 105 | 106 | 107 | 107 |

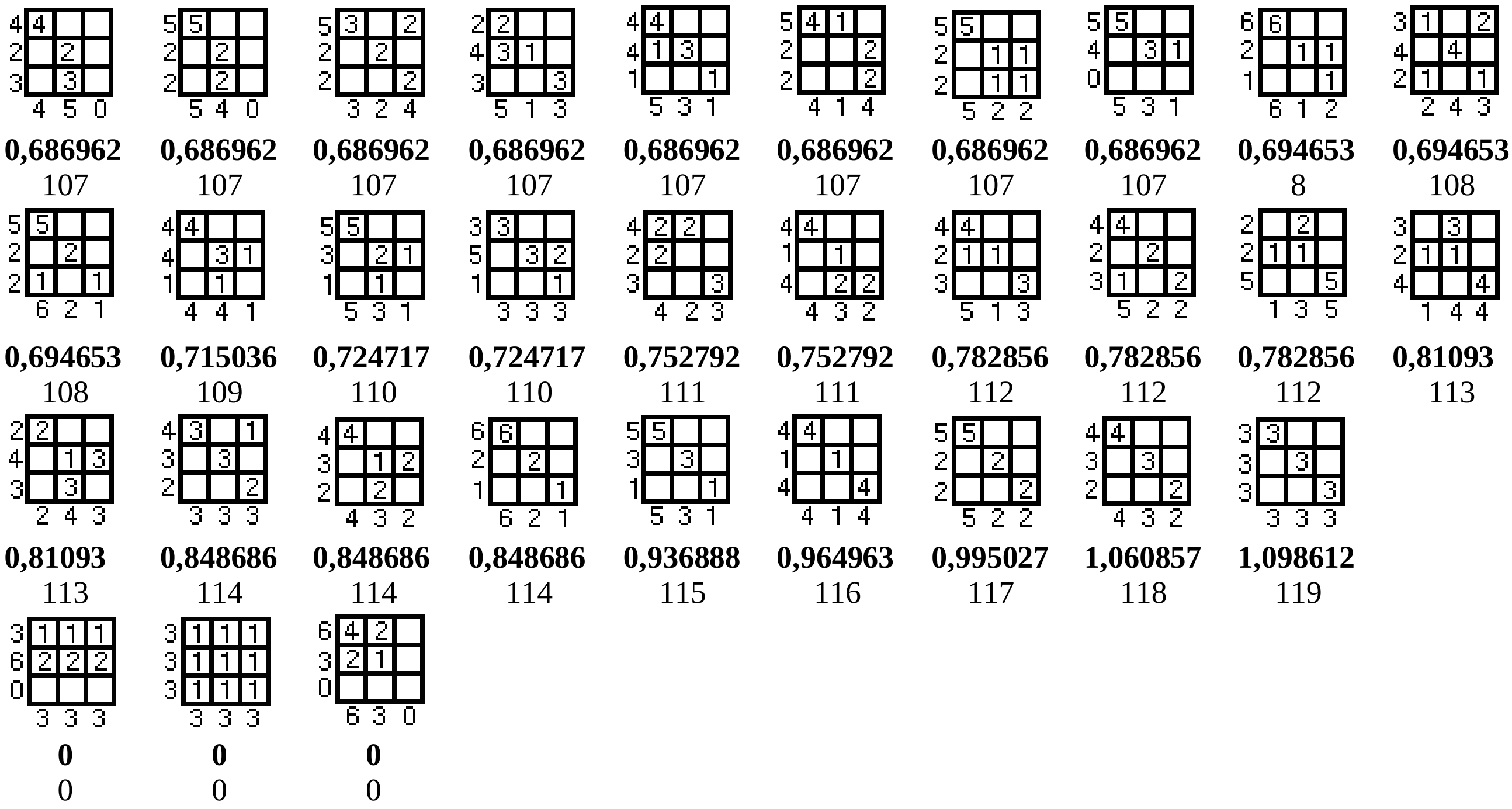

**Figire 1**: Ip values for 3 x 3 matrices with integers from 0 to 8

The variational series of the Ip is well comparable with the variational series of atomic weights of chemical elements. The first and the second members of each series differ by several times, and there can be observed no more differences of such type.

The correlation field of the variational series members is of linear character approximately to the 95th pair of values deviating then from linear relationship. It can be qualitatively compared with only the first naturally occurring 95 chemical elements of the Periodic Table. There can be observed some other similarities between the members of the variational series. The corresponding Ip values of the variational series were attributed to the chemical elements in the order of their position in the Periodic Table (Tab.1).

**Table 1**: Ip values for chemical elements

| Number | Element | Ip | Number | Element | Ip | Number | Element | Ip | Number | Element | Ip |
|---|---|---|---|---|---|---|---|---|---|---|---|
| 1 | **H** | 0.001778 | 31 | **Ga** | 0.155811 | 61 | **Pm** | 0.308065 | **91** | **Pa** | 0.512546 |
| 2 | **He** | 0.005001 | 32 | **Ge** | 0.159033 | 62 | **Sm** | 0.313066 | **92** | **U** | 0.529706 |
| 3 | **Li** | 0.012692 | 33 | **As** | 0.163502 | 63 | **Eu** | 0.317535 | **93** | **Np** | 0.532929 |
| 4 | **Be** | 0.013925 | 34 | **Se** | 0.166725 | 64 | **Gd** | 0.320757 | **94** | **Pu** | 0.54062 |
| 5 | **B** | 0.017161 | 35 | **Br** | 0.171193 | 65 | **Tb** | 0.328449 | **95** | **Am** | 0.548311 |
| 6 | **C** | 0.020383 | 36 | **Kr** | 0.174416 | 66 | **Dy** | 0.33614 | **96** | **Cm** | 0.561003 |
| 7 | **N** | 0.029853 | 37 | **Rb** | 0.182107 | 67 | **Ho** | 0.348832 | **97** | **Bk** | 0.562993 |
| 8 | **O** | 0.039533 | 38 | **Sr** | 0.183885 | 68 | **Er** | 0.350822 | **98** | **Cf** | 0.570684 |
| 9 | **F** | 0.048458 | 39 | **Y** | 0.187108 | 69 | **Tm** | 0.358513 | **99** | **Es** | 0.598759 |
| 10 | **Ne** | 0.050447 | 40 | **Zr** | 0.193566 | 70 | **Yb** | 0.36476 | **100** | **Fm** | 0.621132 |
| 11 | **Na** | 0.056694 | 41 | **Nb** | 0.194799 | 71 | **Lu** | 0.375673 | **101** | **Md** | 0.628823 |
| 12 | **Mg** | 0.058139 | 42 | **Mo** | 0.196789 | 72 | **Hf** | 0.378896 | **102** | **No** | 0.636514 |
| 13 | **Al** | 0.063139 | 43 | **Tc** | 0.20448 | 73 | **Ta** | 0.383365 | **103** | **Lr** | 0.656898 |
| 14 | **Si** | 0.064385 | 44 | **Ru** | 0.210727 | 74 | **W** | 0.386587 | **104** | **Rf** | 0.658887 |
| 15 | **P** | 0.067608 | 45 | **Rh** | 0.212171 | 75 | **Re** | 0.394279 | **105** | **Db** | 0.67427 |
| 16 | **S** | 0.070831 | 46 | **Pd** | 0.221641 | 76 | **Os** | 0.405738 | **106** | **Sg** | 0.683739 |
| 17 | **Cl** | 0.075299 | 47 | **Ag** | 0.224863 | 77 | **Ir** | 0.40896 | **107** | **Bh** | 0.686962 |
| 18 | **Ar** | 0.078522 | 48 | **Cd** | 0.229332 | 78 | **Pt** | 0.416652 | **108** | **Hs** | 0.694653 |
| 19 | **K** | 0.097672 | 49 | **In** | 0.232555 | 79 | **Au** | 0.424343 | **109** | **Mt** | 0.715036 |
| 20 | **Ca** | 0.098905 | 50 | **Sn** | 0.251705 | 80 | **Hg** | 0.433812 | **110** | **Ds** | 0.724717 |
| 21 | **Sc** | 0.100895 | 51 | **Sb** | 0.252938 | 81 | **Tl** | 0.437035 | **111** | **Rg** | 0.752792 |
| 22 | **Ti** | 0.105363 | 52 | **Te** | 0.254928 | 82 | **Pb** | 0.444726 | **112** | **Cn** | 0.782856 |
| 23 | **V** | 0.108586 | 53 | **I** | 0.259396 | 83 | **Bi** | 0.446716 | **113** | **Uut** | 0.81093 |
| 24 | **Cr** | 0.116277 | 54 | **Xe** | 0.262619 | 84 | **Po** | 0.462098 | **114** | **Fl** | 0.848686 |
| 25 | **Mn** | 0.123969 | 55 | **Cs** | 0.27031 | 85 | **At** | 0.465109 | **115** | **Uup** | 0.936888 |
| 26 | **Fe** | 0.125747 | 56 | **Ba** | 0.279779 | 86 | **Rn** | 0.471568 | **116** | **Lv** | 0.964963 |
| 27 | **Co** | 0.128969 | 57 | **La** | 0.283002 | 87 | **Fr** | 0.47479 | **117** | **Uus** | 0.995027 |
| 28 | **Ni** | 0.136661 | 58 | **Ce** | 0.290693 | 88 | **Ra** | 0.482481 | **118** | **Uuo** | 1.060857 |
| 29 | **Cu** | 0.13865 | 59 | **Pr** | 0.292683 | 89 | **Ac** | 0.502865 | 119 | | 1.098612 |
| 30 | **Zn** | 0.146341 | 60 | **Nd** | 0.300374 | 90 | **Th** | 0.504854 | | | |

The similarity of the two variational series allowed to assume that the Ips reflect the structure of atoms nuclei being simulated with the help of integers. This nucleus structure model represents 27 elementary cubes composing a large cube. In such cube integers from 0 to 8 are arranged in such a way that projected on three planes of the Cartesian coordinate system we obtain 3 x 3 matrices described above.

Computer program randomly put numbers into 3 x 3 matrix of (x,y) plane. The numbers were next in the same way randomly distributed in three matrices replacing each other in the direction of Oz axis. Thus the large cube was simulated. Then there were calculated three Ips for 3 x 3 matrices projected on (y,z) and (x,z) planes. The computer program is available on-line www.skyproject.org/atom.aspx.

It was found that 18 chemical elements can be compared with such distributions of integers from 0 to 8 in the large cube, that Ip values for matrices in (x,y), (y,z) and (x,z) projections will be equal. They are helium, beryllium, silicon, germanium, krypton, rhodium, tungsten, iridium and ten chemical elements with diagonal matrices along three axes - holmium, uranium, nobelium, seaborgium, bohrium, flerovium, Uup, livermorium, Uus, Uuo and a hypothetical 119th chemical element. The mentioned states of the nuclear structure are supposed to be called quasiisotope.

If Ip values projected on two planes are equal, such structure of atom nucleus in many cases corresponds to an isotope of the chemical element. This is the basic, dominating, yet not the only type of nucleus structure, determining an isotope existence. There were not found any isotope states of atom nuclei of Pm, Yb, Ta, Re, Os, Pt, Hg, Bi, Po, At, Rn, Th, Pa, U, Pu, Am, Cm, Bk and all chemical elements starting with number 99. Isotope states of atom nuclei for chemical elements and frequencies of their occurrence at random selection of all possible states are given in Table 2. For every chemical element there were generated from 500 thousand to 2.5 million random spatial combinations of numbers.

**Table 2**: Frequencies of isotope states of chemical elements

| Elements | Frequencies of isotope states | | | | | | | | | | | | | | | | | | | | | | |
|---|---|---|---|---|---|---|---|---|---|---|---|---|---|---|---|---|---|---|---|---|---|---|---|
| H | **Si** | **Ge** | **Ir** | **U** | **Bh** | | | | | | | | | | | | | | | | | | |
| | 725 | 7 | 5 | 127 | 241 | | | | | | | | | | | | | | | | | | |
| He | **Al** | **Ge** | **Sm** | **Ir** | **Fm** | **Bh** | **Uus** | | | | | | | | | | | | | | | | |
| | 74 | 171 | 122 | 55 | 10 | 116 | 1 | | | | | | | | | | | | | | | | |
| Li | **He** | **Al** | **Cr** | **Mn** | **Ge** | **Kr** | **W** | **Ir** | **Au** | **Pu** | **Es** | **No** | **Bh** | **Rg** | **Uut** | **Fl** | **Uuo** | | | | | | |
| | 194 | 244 | 862 | 67 | 132 | 447 | 232 | 250 | 51 | 26 | 12 | 61 | 117 | 6 | 2 | 2 | 1 | | | | | | |
| Be | **Ho** | | | | | | | | | | | | | | | | | | | | | | |
| | 203 | | | | | | | | | | | | | | | | | | | | | | |
| B | **Si** | **Mn** | **U** | **No** | | | | | | | | | | | | | | | | | | | |
| | 747 | 42 | 325 | 314 | | | | | | | | | | | | | | | | | | | |
| C | **Kr** | **W** | **Es** | **No** | **Uut** | **Fl** | **Uuo** | | | | | | | | | | | | | | | | |
| | 211 | 9 | 5 | 168 | 1 | 6 | 5 | | | | | | | | | | | | | | | | |
| N | **Be** | **Na** | **Si** | **Ca** | **Ru** | **Sb** | **Ho** | **Yb** | **At** | **U** | | | | | | | | | | | | | |
| | 165 | 77 | 98 | 58 | 34 | 2 | 108 | 9 | 2 | 29 | | | | | | | | | | | | | |
| O | **Si** | **Mo** | **Ir** | **Bi** | **U** | **Rf** | **Ds** | **Uup** | | | | | | | | | | | | | | | |
| | 686 | 24 | 4 | 27 | 327 | 1 | 5 | 2 | | | | | | | | | | | | | | | |
| F | **Be** | **Mg** | **Cr** | **Mn** | **Rb** | **Cs** | **Dy** | **Ho** | **Re** | **Hs** | | | | | | | | | | | | | |
| | 268 | 342 | 41 | 37 | 32 | 130 | 25 | 185 | 4 | 1 | | | | | | | | | | | | | |
| Ne | **He** | **Sc** | **Mn** | **Cu** | **Ge** | **Kr** | **Mo** | **Er** | **Ir** | **Bi** | **Bk** | **No** | **Ds** | **Uus** | | | | | | | | | |
| | 45 | 40 | 53 | 60 | 4 | 19 | 8 | 36 | 22 | 6 | 1 | 50 | 7 | 1 | | | | | | | | | |
| Na | **U** | | | | | | | | | | | | | | | | | | | | | | |
| | 117 | | | | | | | | | | | | | | | | | | | | | | |
| Mg | **Mn** | **Kr** | **Au** | **Am** | **No** | **Fl** | | | | | | | | | | | | | | | | | |
| | 94 | 160 | 7 | 9 | 179 | 5 | | | | | | | | | | | | | | | | | |
| Al | **He** | **Ge** | **Bh** | | | | | | | | | | | | | | | | | | | | |
| | 77 | 26 | 38 | | | | | | | | | | | | | | | | | | | | |
| Si | **Na** | **Ru** | **Yb** | **U** | **Sg** | | | | | | | | | | | | | | | | | | |
| | 289 | 291 | 61 | 88 | 12 | | | | | | | | | | | | | | | | | | |
| P | **Si** | **In** | **Pb** | **U** | **Lr** | | | | | | | | | | | | | | | | | | |
| | 101 | 9 | 6 | 58 | 1 | | | | | | | | | | | | | | | | | | |

| **Elements** | **Frequencies of isotope states** | | | | | | | | | | | | | | | | | | | | | | |
|---|---|---|---|---|---|---|---|---|---|---|---|---|---|---|---|---|---|---|---|---|---|---|---|
| **S** | **He** | **Be** | **F** | **Al** | **Ca** | **Sc** | **Cu** | **Ge** | **Kr** | **Mo** | **Te** | **Ce** | **Pr** | **Sm** | **Ho** | **Er** | **W** | **Ir** | **Au** | **At** | **Ac** | **Pu** | **Am** |
| | 150 | 1133 | 187 | 101 | 864 | 1252 | 75 | 788 | 130 | 763 | 157 | 16 | 5 | 130 | 543 | 100 | 33 | 143 | 9 | 16 | 7 | 1 | 7 |
| | **Es** | **Fm** | **Bh** | **Mt** | **Fl** | **Uus** | **Uuo** | | | | | | | | | | | | | | | | |
| | 2 | 1 | 110 | 9 | 8 | 2 | 1 | | | | | | | | | | | | | | | | |
| **Cl** | **Si** | **U** | **Am** | **Fl** | | | | | | | | | | | | | | | | | | | |
| | 8 | 196 | 6 | 4 | | | | | | | | | | | | | | | | | | | |
| **Ar** | **Mn** | **Kr** | **Nb** | **Sb** | **At** | **No** | **Mt** | **Rg** | **Uut** | **Lv** | | | | | | | | | | | | | |
| | 89 | 39 | 13 | 9 | 8 | 11 | 7 | 2 | 1 | 2 | | | | | | | | | | | | | |
| **K** | **Si** | **U** | **Bk** | | | | | | | | | | | | | | | | | | | | |
| | 187 | 204 | 2 | | | | | | | | | | | | | | | | | | | | |
| **Ca** | **He** | **Li** | **Be** | **Ar** | **Co** | **Ni** | **Ge** | **Se** | **Y** | **Nb** | **Ag** | **In** | **Sb** | **La** | **Ce** | **Ho** | **Hf** | **W** | **Tl** | **Ac** | | | |
| | 36 | 276 | 1534 | 256 | 223 | 528 | 55 | 135 | 37 | 110 | 50 | 213 | 309 | 59 | 207 | 527 | 47 | 137 | 41 | 131 | | | |
| **Sc** | **Rf** | **Bh** | **Ds** | **Uup** | | | | | | | | | | | | | | | | | | | |
| | 8 | 26 | 1 | 4 | | | | | | | | | | | | | | | | | | | |
| **Ti** | **Na** | **Mg** | **Si** | **Cr** | **Mn** | **Rb** | **Ru** | **Rh** | **Cs** | **Pm** | **Dy** | **Yb** | **Re** | **U** | **Sg** | **Hs** | | | | | | | |
| | 96 | 26 | 113 | 20 | 5 | 47 | 158 | 142 | 10 | 1 | 2 | 24 | 20 | 61 | 8 | 1 | | | | | | | |
| **V** | **Sc** | **Mn** | **Cu** | **Ge** | **Kr** | **Mo** | **Te** | **Tb** | **Er** | **W** | **Ir** | **Au** | **Bi** | **Pu** | **Am** | **Bk** | **Es** | **No** | **Rf** | **Bh** | **Ds** | **Fl** | **Uup** |
| | 27 | 110 | 23 | 21 | 188 | 43 | 373 | 29 | 23 | 32 | 22 | 37 | 33 | 13 | 57 | 20 | 19 | 128 | 9 | 169 | 6 | 19 | 5 |
| **Cr** | **Mn** | **Kr** | **W** | **Au** | **Pb** | **Am** | **No** | **Lr** | **Hs** | **Fl** | **119** | | | | | | | | | | | | |
| | 280 | 105 | 4 | 2 | 17 | 15 | 89 | 1 | 1 | 6 | 1 | | | | | | | | | | | | |
| **Mn** | **No** | | | | | | | | | | | | | | | | | | | | | | |
| | 17 | | | | | | | | | | | | | | | | | | | | | | |
| **Fe** | **Si** | **U** | | | | | | | | | | | | | | | | | | | | | |
| | 111 | 485 | | | | | | | | | | | | | | | | | | | | | |
| **Co** | **He** | **Ge** | **Bh** | **Mt** | **Lv** | | | | | | | | | | | | | | | | | | |
| | 75 | 111 | 14 | 14 | 4 | | | | | | | | | | | | | | | | | | |
| **Ni** | **Be** | **Ca** | **Mn** | **Zn** | **Kr** | **Nb** | **Rh** | **Sb** | **Cs** | **Ho** | **W** | **Au** | **At** | **Am** | **Es** | **No** | **Mt** | **Rg** | **Uut** | **Fl** | **Lv** | **Uuo** | |
| | 5715 | 217 | 579 | 504 | 644 | 11 | 672 | 10 | 985 | 2857 | 84 | 767 | 7 | 21 | 27 | 36 | 4 | 5 | 3 | 4 | 3 | 2 | |
| **Cu** | **Er** | **Bi** | **Rf** | **Ds** | | | | | | | | | | | | | | | | | | | |
| | 25 | 37 | 2 | 9 | | | | | | | | | | | | | | | | | | | |
| **Zn** | **No** | **Rf** | **Db** | **Uup** | **119** | | | | | | | | | | | | | | | | | | |
| | 82 | 2 | 2 | 12 | 1 | | | | | | | | | | | | | | | | | | |
| | | | | | | | | | | | | | | | | | | | | | | | |

| Elements | Frequencies of isotope states | | | | | | | | | | | | | | | | | | | | | |
|---|---|---|---|---|---|---|---|---|---|---|---|---|---|---|---|---|---|---|---|---|---|---|
| **Ga** | **He** | **Na** | **Al** | **Si** | **Sc** | **Ge** | **Mo** | **Ru** | **Te** | **Pr** | **Sm** | **Er** | **Yb** | **Ir** | **U** | **Sg** | | | | | | |
| | 53 | 238 | 5 | 28 | 95 | 16 | 27 | 238 | 10 | 9 | 26 | 12 | 56 | 16 | 69 | 10 | | | | | | |
| **Ge** | **He** | **Bk** | **Bh** | | | | | | | | | | | | | | | | | | | |
| | 58 | 3 | 54 | | | | | | | | | | | | | | | | | | | |
| **As** | **Si** | **U** | | | | | | | | | | | | | | | | | | | | |
| | 71 | 75 | | | | | | | | | | | | | | | | | | | | |
| **Se** | **He** | **Ge** | **Kr** | **Te** | **Sm** | **Tb** | **Ho** | **W** | **Au** | **At** | **Ac** | **Pu** | **Am** | **Bk** | **Fm** | **No** | **Lr** | **Rf** | **Bh** | **Mt** | **Uup** | **Uuh** | |
| | 23 | 24 | 27 | 1 | 47 | 5 | 2 | 12 | 3 | 7 | 6 | 2 | 1 | 1 | 2 | 4 | 1 | 1 | 20 | 1 | 2 | 1 | |
| **Br** | Si | **Mn** | | | | | | | | | | | | | | | | | | | | |
| | 36 | 32 | | | | | | | | | | | | | | | | | | | | |
| **Kr** | **C** | **Mg** | **Cr** | **Mn** | **Rb** | **Rh** | **In** | **W** | **Re** | **Au** | **Pb** | **Ra** | **Am** | **No** | **Lr** | **Hs** | **Fl** | | | | | |
| | 210 | 17 | 66 | 86 | 134 | 75 | 174 | 93 | 49 | 9 | 15 | 9 | 7 | 53 | 3 | 8 | 2 | | | | | |
| **Rb** | **No** | **Fl** | | | | | | | | | | | | | | | | | | | | |
| | 259 | 46 | | | | | | | | | | | | | | | | | | | | |
| **Sr** | **Sb** | | | | | | | | | | | | | | | | | | | | | |
| | 16 | | | | | | | | | | | | | | | | | | | | | |
| **Y** | **Bh** | **Mt** | | | | | | | | | | | | | | | | | | | | |
| | 9 | 1 | | | | | | | | | | | | | | | | | | | | |
| **Zr** | **Si** | **Mo** | **Bi** | **Ds** | | | | | | | | | | | | | | | | | | |
| | 36 | 8 | 13 | 3 | | | | | | | | | | | | | | | | | | |
| **Nb** | **Ho** | **Mt** | | | | | | | | | | | | | | | | | | | | |
| | 1770 | 1 | | | | | | | | | | | | | | | | | | | | |
| **Mo** | **Ge** | **Er** | **Ir** | **Bi** | **Bk** | **Rf** | **Bh** | **Ds** | **Uup** | | | | | | | | | | | | | |
| | 39 | 18 | 21 | 23 | 3 | 6 | 16 | 11 | 5 | | | | | | | | | | | | | |
| **Tc** | **Fl** | | | | | | | | | | | | | | | | | | | | | |
| | 4 | | | | | | | | | | | | | | | | | | | | | |
| **Ru** | **U** | | | | | | | | | | | | | | | | | | | | | |
| | 6 | | | | | | | | | | | | | | | | | | | | | |
| **Rh** | **Mg** | **Cr** | **Mn** | **Kr** | **Po** | **No** | **Db** | | | | | | | | | | | | | | | |
| | 91 | 21 | 24 | 171 | 9 | 89 | 1 | | | | | | | | | | | | | | | |
| **Pd** | **Na** | **Ge** | **Y** | **Ru** | **Ag** | **In** | **La** | **Ce** | **Yb** | **Tl** | **Pb** | **U** | **Sg** | | | | | | | | | |
| | 319 | 18 | 11 | 311 | 20 | 14 | 25 | 2 | 77 | 33 | 2 | 205 | 25 | | | | | | | | | |
| | | | | | | | | | | | | | | | | | | | | | | |

| Elements | Frequencies of isotope states | | | | | | | | | | | | | | | | | | | | | | |
|---|---|---|---|---|---|---|---|---|---|---|---|---|---|---|---|---|---|---|---|---|---|---|---|
| **Ag** | **In** | **Ir** | **Rf** | **Bh** | **Hs** | **Ds** | **Uup** | **Lv** | | | | | | | | | | | | | | | |
| | 19 | 110 | 6 | 41 | 2 | 6 | 5 | 3 | | | | | | | | | | | | | | | |
| **Cd** | **Si** | **U** | **Am** | **Fl** | | | | | | | | | | | | | | | | | | | |
| | 79 | 70 | 7 | 4 | | | | | | | | | | | | | | | | | | | |
| **In** | **Kr** | **W** | **Au** | **Am** | **Es** | **No** | **Lg** | **Rg** | **Uut** | **Fl** | **Uuo** | | | | | | | | | | | | |
| | 166 | 19 | 21 | 48 | 45 | 43 | 2 | 2 | 6 | 39 | 6 | | | | | | | | | | | | |
| **Sn** | **Si** | **U** | **Bk** | | | | | | | | | | | | | | | | | | | | |
| | 56 | 56 | 2 | | | | | | | | | | | | | | | | | | | | |
| **Sb** | **Mt** | **Lv** | | | | | | | | | | | | | | | | | | | | | |
| | 11 | 7 | | | | | | | | | | | | | | | | | | | | | |
| **Te** | **Ge** | **Bk** | **Fm** | **Bh** | | | | | | | | | | | | | | | | | | | |
| | 152 | 13 | 2 | 214 | | | | | | | | | | | | | | | | | | | |
| **I** | **Pm** | | | | | | | | | | | | | | | | | | | | | | |
| | 3 | | | | | | | | | | | | | | | | | | | | | | |
| **Xe** | **C** | **Mg** | **Cr** | **Mn** | **Cu** | **Ge** | **Kr** | **Rb** | **Mo** | **Rh** | **In** | **Te** | **Pr** | **Er** | **W** | **Re** | **Ir** | **Au** | **Pb** | **Bi** | **Ra** | **Pu** | **Es** |
| | 36 | 50 | 15 | 6 | 4 | 10 | 136 | 112 | 20 | 2 | 18 | 7 | 15 | 25 | 43 | 120 | 8 | 6 | 9 | 2 | 5 | 5 | 8 |
| | **Lr** | **Rf** | **Bh** | **Hs** | **Ds** | **Uup** | **Uus** | | | | | | | | | | | | | | | | |
| | 4 | 4 | 17 | 11 | 2 | 6 | 2 | | | | | | | | | | | | | | | | |
| **Cs** | **Kr** | **No** | **Fl** | **Uuo** | | | | | | | | | | | | | | | | | | | |
| | 17 | 47 | 3 | 1 | | | | | | | | | | | | | | | | | | | |
| **Ba** | **U** | | | | | | | | | | | | | | | | | | | | | | |
| | 54 | | | | | | | | | | | | | | | | | | | | | | |
| **La** | **Fm** | **Bh** | **Mt** | **Lv** | **Uus** | | | | | | | | | | | | | | | | | | |
| | 12 | 138 | 2 | 4 | 3 | | | | | | | | | | | | | | | | | | |
| **Ce** | **Cr** | **Kr** | **In** | **Sb** | **At** | **Mt** | **Uut** | **Lv** | **Uuo** | | | | | | | | | | | | | | |
| | 8 | 32 | 35 | 24 | 10 | 21 | 6 | 2 | 2 | | | | | | | | | | | | | | |
| **Pr** | **Uup** | | | | | | | | | | | | | | | | | | | | | | |
| | 2 | | | | | | | | | | | | | | | | | | | | | | |
| **Nd** | **Db** | | | | | | | | | | | | | | | | | | | | | | |
| | 1 | | | | | | | | | | | | | | | | | | | | | | |
| **Sm** | **He** | **Ge** | **Mo** | **Ir** | **Bh** | | | | | | | | | | | | | | | | | | |
| | 16 | 96 | 52 | 13 | 4 | | | | | | | | | | | | | | | | | | |
| **Eu** | | | | | | | | | | | | | | | | | | | | | | | |

| Elements | Frequencies of isotope states | | | | | | | | | | | | | |
|---|---|---|---|---|---|---|---|---|---|---|---|---|---|---|
| | **Au** | **U** | **Sg** | | | | | | | | | | | |
| | 129 | 285 | 37 | | | | | | | | | | | |
| **Gd** | **Cr** | **Kr** | **Mo** | **In** | **W** | **Pu** | **Am** | **Bk** | **Es** | **Hs** | **Ds** | **Fl** | **Uup** | **Uuo** |
| | 18 | 25 | 57 | 22 | 55 | 3 | 14 | 2 | 17 | 8 | 28 | 9 | 2 | 2 |
| **Tb** | **Mn** | **Sb** | **Pb** | **No** | **Lr** | **Hs** | **Fl** | | | | | | | |
| | 268 | 213 | 36 | 12 | 3 | 3 | 44 | | | | | | | |
| **Dy** | **No** | | | | | | | | | | | | | |
| | 23 | | | | | | | | | | | | | |
| **Ho** | **Ni** | **Fl** | **Lv** | **Uuo** | | | | | | | | | | |
| | 571 | 10 | 13 | 2 | | | | | | | | | | |
| **Er** | **Cu** | **Mo** | **Bi** | **Rf** | **Ds** | **Uus** | | | | | | | | |
| | 49 | 53 | 67 | 8 | 8 | 8 | | | | | | | | |
| **Tm** | **No** | | | | | | | | | | | | | |
| | 4 | | | | | | | | | | | | | |
| **Lu** | **U** | | | | | | | | | | | | | |
| | 751 | | | | | | | | | | | | | |
| **Hf** | **Ge** | **Mo** | **In** | **At** | **Ac** | **Bh** | **Mt** | **Uuo** | | | | | | |
| | 165 | 51 | 49 | 29 | 13 | 4 | 2 | 4 | | | | | | |
| **W** | **Cr** | **Kr** | **Hf** | **Tl** | **Ac** | **Am** | **No** | **Hs** | **Fl** | | | | | |
| | 19 | 29 | 6 | 33 | 58 | 11 | 11 | 25 | 4 | | | | | |
| **Ir** | **Cu** | **Mo** | **Te** | **Sm** | **Er** | **Bk** | **Rf** | **Bh** | | | | | | |
| | 7 | 6 | 8 | 38 | 10 | 2 | 2 | 5 | | | | | | |
| **Au** | **No** | **119** | | | | | | | | | | | | |
| | 61 | 1 | | | | | | | | | | | | |
| **Tl** | **Sm** | **Bh** | **Mt** | **Lv** | | | | | | | | | | |
| | 28 | 51 | 3 | 2 | | | | | | | | | | |
| **Pb** | **Fl** | **Uuo** | | | | | | | | | | | | |
| | 2 | 3 | | | | | | | | | | | | |
| **Fr** | **Kr** | **Mo** | **Ir** | **Lr** | **Rf** | **Bh** | **Hs** | **Ds** | **Uus** | | | | | |
| | 7 | 3 | 29 | 1 | 4 | 21 | 1 | 3 | 1 | | | | | |
| **Ra** | **No** | **Fl** | **Uuo** | | | | | | | | | | | |
| | 518 | 53 | 6 | | | | | | | | | | | |
| | | | | | | | | | | | | | | |

| **Elements** | **Frequencies of isotope states** | | | | | | | | | | | | | | | | | | | | | | |
|---|---|---|---|---|---|---|---|---|---|---|---|---|---|---|---|---|---|---|---|---|---|---|---|
| **Ac** | **Mt** | **Lv** | | | | | | | | | | | | | | | | | | | | | |
| | 8 | 2 | | | | | | | | | | | | | | | | | | | | | |
| **Np** | **Bh** | | | | | | | | | | | | | | | | | | | | | | |
| | 245 | | | | | | | | | | | | | | | | | | | | | | |
| **Cf** | **Uup** | | | | | | | | | | | | | | | | | | | | | | |
| | 10 | | | | | | | | | | | | | | | | | | | | | | |

The isotope states of atom nucleus is favorable for the spatial association of chemical elements, and appears as the property of their being together in nature. The property is supposed to be named as nucleus affinity. It can be evident in the existing of many chemical elements in native form that can be explained by natural nucleus affinity of chemical element to itself due to atom nucleus structure. Chemical elements being abundant in nuggets show nucleus affinity to small number of elements.

Another example – terbium can be characterized by affinity to manganese; therefore terbium is present in iron-manganese concretions of the World Ocean. The composition of the Earth lithosphere can be considered as a result of H, B, N, O, P, Cl, K, Fe, Ga, As, Br, Zr, Cd, Sn to Si, and Si to Na nucleus affinity. It seems to be more difficult to explain the presence of high concentrations of Mg, Ca, Al and C and some other chemical elements having no affinity to Si in the lithosphere. It can be done by studying the affinity of these elements to the others typical for the lithosphere.

Co-existence of chemical elements of five-element formation Ni-Co- Ag –As (U)- Bi in nature in industrial concentrations can be explained with nucleus affinity of the first three elements to calcium, silver to iridium, bismuth and uranium to oxygen being affined to iridium. Arsenic possesses affinity to uranium. Thus, elements with different chemical and physical properties can coexist in nature.

The value of nucleus affinity for a chemical element to another is proposed to determine by the number of corresponding isotope states obtained as a result of random sorting of variants of atom nucleus structure. For example, affinity for terbium to Mn is stronger than to Sb (see Table 2).

Isotope states of nucleus structure of chemical elements are compared with the isotopes of elements themselves. It is well visible for light chemical elements. For hydrogen and helium the number of such states to stable chemical elements is equal to the number of stable and radioactive isotopes. The regularity is not observed for lithium possessing nucleus affinity to many stable chemical elements. Beryllium with one stable isotope can be characterized by nucleus affinity only to holmium.

For boron it is pointed out the deficiency of two isotope states for the explanation of two out of four radioactive isotopes - $^{8}B$ and $^{9}B$. Here is another rule of isotope states formation, presumably with different Ip for all three planes of the coordinate system. For both given chemical element and oxygen, two isotope states to stable chemical elements are observed. They correspond to the number of stable isotopes of these elements.

The above given regularities are broken for heavier chemical elements, while important coincidences are noted for chemical elements with only one isotope. The atoms of As, Nb, Cs and Ho each possess one isotope state to stable chemical elements. Na, I, Pr, Tm and Au each are characterized with this state to a radioactive element.

The data obtained makes it possible to explain why isotopes are found in nature in different percentage ratios to each other. It can be related to the probability of selecting isotope states of chemical elements in random selection of all possible variants of atoms nuclei structure. Herewith for H, He and C the isotope ratios obtained by computer simulation must be close to the theoretical ones.

Breaking precise theoretical ratios in nature lets us assume the idea of selective synthesis of isotopes after the Big Bang. The synthesis of helium isotopes in the conditions of big energies resulted in the selective choice of predominantly $^{4}He$ isotope compared with $^{3}He$ (progressive selective synthesis). Later in the conditions of lower energies while forming $^{1}H$ and $^{2}H$ isotopes there was more likely synthesized selectively $^{1}H$ isotope (regressive selective synthesis).

The results of computer simulation show high potential of nucleus structure research for the characterization of their chemical and physical properties.

**ACKNOWLEDGEMENTS**

I express gratitude to my wife Elena Fomina for patience and support as well as the translation of the paper in English. I am also grateful to Timothy Labushev for coding the computer program of simulation of atom nucleus structure. I would like to acknowledge Microsoft Corporation for outstanding software.